\documentclass[american,english,prx]{revtex4-2}
\usepackage[T1]{fontenc}
\usepackage[utf8]{inputenc}
\usepackage{amsmath}
\usepackage{amssymb}
\usepackage{graphicx}
\usepackage{esint}
\usepackage{babel}
\begin{document}
\global\long\def\ket#1{\left|#1\right\rangle }%

\global\long\def\bra#1{\left\langle #1\right|}%

\global\long\def\braket#1#2{\left\langle #1\left|#2\right.\right\rangle }%

\global\long\def\ketbra#1#2{\left|#1\right\rangle \left\langle #2\right|}%

\global\long\def\braOket#1#2#3{\left\langle #1\left|#2\right|#3\right\rangle }%

\global\long\def\mc#1{\mathcal{#1}}%

\global\long\def\nrm#1{\left\Vert #1\right\Vert }%

\title{Orders of magnitude sampling overhead reduction in quantum error mitigation }
\author{Raam Uzdin}
\affiliation{\selectlanguage{american}%
Fritz Haber Research Center for Molecular Dynamics, Institute of Chemistry,
The Hebrew University of Jerusalem, Jerusalem 9190401, Israel}
\selectlanguage{english}%
\begin{abstract}
Quantum error mitigation (QEM) infers noiseless expectation values
from noisy variants of a target quantum circuit. Unlike quantum error
correction, QEM requires no additional hardware resources and is therefore
routinely employed in experiments on contemporary quantum processors.
QEM strategies based on agnostic noise amplification (ANA) are intrinsically
resilient to temporal noise drift during the execution of the experiment,
but their sampling cost (runtime overhead) remains a major practical
bottleneck. In this work, we introduce the virtual noise scaling framework
and combine it with layered mitigation to further enhance performance.
While virtual noise scaling consistently reduces sampling overhead,
we identify a specific noise threshold for the layered mitigation
approach. When the noise level is above this threshold, layered mitigation
decreases the sampling overhead; conversely, when below it, the overhead
increases. Notably, this threshold is circuit-independent and depends
solely on the number of layers. For strong noise, the combination
of virtual noise scaling and layered mitigation yields several orders
of magnitude reduction in sampling overhead compared with conventional
zero-noise extrapolation post-processing. As a result, mitigation
tasks that once seemed unrealistic are now challenging but achievable.
The proposed approach is compatible with dynamic circuits and can
be seamlessly integrated with error detection and quantum error correction
schemes. In addition, it is also applicable to ANA-based mitigation
of mid-circuit measurements and preparation errors. Our findings extend
to any constant-step amplification factors and therefore also apply
to probabilistic error amplification (PEA) QEM. We validate our post-processing
approach by applying it to previously reported experimental data,
where we observe a substantial improvement in mitigation efficiency
and accuracy.
\end{abstract}
\maketitle
Even after error correction is successfully implemented, quantum computers
will not be noise-free, as quantum error correction schemes (QEC)
struggle with correlated errors, leakage errors, and other non-idealities.
An emerging strategy \foreignlanguage{american}{\cite{suzuki2022quantum,2021errorQEMforTgatesQEC,lostaglio2021error,Qedma2025ImportanceConcept,wahl2023ZNEforQEC,Qedma2025syndrome,QECQEMsoftInfo,Cai2025virtualChan,ZNEfitLogicalQubits2026}}
is to complement QEC with quantum error mitigation (QEM) protocols
\foreignlanguage{american}{\cite{cai2022quantum,qin2022overviewQEM,endo2021hybrid,temme2017ZNEPECorig,quek2024exponentially,takagi2022fundamental,li2017efficient,endo2018practical,strikis2021learning,czarnik2021error,koczor2021exponential,huggins2021virtual,Tiron,cai2021symmetry,mari2021extending,huo2022DualState,lowe2021unified,nation2021scalable,liao2024MLqem,bravyi2021mitigating,kim2021scalable,berg2022probabilistic,kandala2019error,song2019quantum,arute2020hartree,urbanek2021mitigating,zhang2020error,sagastizabal2019experimental,he2020zero,TEMqem,kim2020neural,cantori2024synergy,araki2025spacetimeVD,bako2025exponentialDualState}}
which are already widely used for handling errors in the pre-fault-tolerant
era \foreignlanguage{american}{\cite{huggins2021virtual,kim2021scalable,kandala2019error,song2019quantum,arute2020hartree,urbanek2021mitigating,zhang2020error,sagastizabal2019experimental,Quantinuum2025QuantumMagnetismPEA,kim2023evidence,kim2024IBMstabilizeNoisePEC,yang2022sparseReadout}}.
In \cite{Cai2026EMPhysQQEC} it was shown that any linear QEM can
be seamlessly implemented before QEC without affecting the AEC protocol.
QEM methods remove the noise bias from the expectation values at the
cost of increasing the variance, thereby increasing the sampling cost.
To counter this increase, more shots have to be executed and the runtime
increases accordingly. Crucially, unlike QEC, QEM involves little
or no hardware overhead.

The two most widely used schemes are probabilistic error cancellation
(PEC) and agnostic noise amplification (ANA). In PEC and its variants
\foreignlanguage{american}{\cite{temme2017ZNEPECorig,mari2021extending,PECtimeDrift}},
noise characterization and a simplified sparse noise model are required
for removing the noise from expectation values. As such, these methods
are vulnerable to noise drifts in which the noise parameters change
during the execution of the experiment. As a result, the learned noise
model no longer represents the actual noise and residual noise appears.
ANA, on the other hand, does not involve characterization. In this
approach, the noise is amplified by a known factor by constructing
special noisy variants of the target circuit. The ANA approach has
two key advantages (1) it can mitigate non-sparse noise models which
are very difficult to address in PEC, (2) it can be made fully resilient
to temporal drifts of the noise parameters, i.e. handle non-stationary
noise. Note that the widely used zero noise extrapolation (ZNE) is
not necessarily noise agnostic. ZNE is a post-processing technique
that uses measurement with multiple noise amplification factors to
predict the noiseless expectation values but the noise amplification
can be done either in an agnostic manner or using characterization
as done in probabilistic error amplification (PEA) \cite{kim2023evidence}.
We point out that a class of methods called virtual purification \foreignlanguage{american}{\cite{Koczor2021,huggins2021virtual,araki2025spacetimeVD}}
is also agnostic but it is applicable only to special types of noise.
Furthermore, the purification approach involves hardware overhead.

To experimentally implement ANA, various techniques have been proposed,
including (i) pulse stretching \foreignlanguage{american}{\cite{temme2017ZNEPECorig,kim2023scalable}}
(ii) digital local folding, also known as gate insertion, and digital
global folding\foreignlanguage{american}{ \cite{majumdar2023bestPracticeDZNE,PhysRevA.105.042406,PhysRevA.102.012426,schultz2022impact}},
and (iii) the pulse-inverse KIK \cite{npjqiKIK} method and its refined
and enhanced version ``Layered KIK'' \cite{LKIK}. Pulse stretching
is challenging to implement and calibrate and is also inconsistent
with twirling techniques \cite{knill2004PauliTwirl,Emerson2020LearnTwirlExpIBM,cai2019SmallerTwirl,2024PST}
for treating coherent errors. Digital folding has been shown analytically
and experimentally to be incorrect even at the leading order of the
noise, leading to results which significantly deviate from the ideal
value (regardless of the sampling overhead invested). 

In this work we start by studying the fidelity limit of the Taylor-based
post-processing and introduce a method for reducing the runtime overhead
by orders of magnitude when the noise is strong. 

\section{Preliminaries}

\subsection{Noise amplification in Liouville space}

In this work we describe the quantum state of the system using density
vectors $\ket{\rho}$ in Liouville space \cite{gyamfi2020fundamentals}.
The vector $\ket{\rho}$ is obtained by flattening the density matrix
into a column vector. In this representation, quantum channels act
linearly from the left on the state, which makes it convenient for
describing noisy quantum dynamics.

If the ideal evolution in Hilbert space is given by a unitary operator
$U_{H}$, then the corresponding ideal evolution in Liouville space
is $U=U_{H}\otimes U_{H}^{*}$. The noisy evolution operator can be
written as
\begin{equation}
K=UN.
\end{equation}
where $N$ is the noise channel operator in Liouville space. Noise
amplification by a factor $\alpha$ is defined as
\begin{equation}
K_{amp}=UN^{\alpha}.
\end{equation}

In this work we focus on odd amplification powers. Excluding pulse
stretching, all agnostic noise amplification (ANA) methods naturally
lead to odd amplification factors. This originates from the use of
a circuit (or gate) together with its inverse in order to amplify
the noise. As shown in \cite{LKIK}, if the evolution operator is
sliced into layers $K=K_{\ensuremath{1}}\dots K_{L}$ then, when the
number of layers is sufficiently large, the following relation holds:
\begin{equation}
UN^{2j+1}\cong K_{1}(K_{1}^{I}K_{1})^{j}\dots K_{L}(K_{L}^{I}K_{L})^{j},
\end{equation}
where $K_{\ell}^{I}$ denotes the pulse inverse of layer $\ell$.
Throughout this work, the symbol $\cong$ is used to denote equality
for all practical purposes, meaning that the approximation becomes
arbitrarily accurate when a sufficiently large number of layers is
used. We therefore assume that, for practical implementations, perfect
noise amplification can be achieved.

Finally, we note that even if the native noise generators, that is,
the Lindbladians, are Hermitian, the effective noise channel N is
generally not Hermitian due to time ordering. In a Magnus expansion,
$N=e^{\sum_{k}\Omega_{k}}$, the leading term $\Omega_{1}$ is Hermitian
when the Lindbladians are Hermitian, whereas the second-order term
$\Omega_{2}$ is anti-Hermitian. Noise Hermiticity plays an important
role in the present work. In the next section, we show how Hermiticity
can be effectively restored, even when the given noise channel is
not Hermitian. 

\subsection{Taylor-based mitigation}

Using $K=UN$ for the noisy circuit evolution operator, the Taylor-based
$m$-th order mitigated evolution operator is \cite{npjqiKIK}
\begin{align}
K_{mit}^{(m)} & \cong\sum_{k=0}^{m}a_{j,Tay}^{(m)}UN^{2j+1},\\
a_{j,Tay}^{(m)} & =\frac{(-1)^{j}(2m+1)!!}{2^{m}(2j+1)j!(m-j)!}=\frac{(-1)^{j}}{2j+1}\left(\begin{array}{c}
m+\frac{1}{2}\\
m
\end{array}\right)\left(\begin{array}{c}
m\\
j
\end{array}\right),
\end{align}
Where $\left(\begin{array}{c}
x\\
y
\end{array}\right)$ is the Binomial function. The Taylor coefficients $a_{j,}{}_{Tay}^{(m)}$
coincide with the Richardson extrapolation coefficients \cite{2021errorQEMforTgatesQEC,Tiron}
used in zero-noise extrapolation (ZNE) when odd noise-amplification
factors are employed \cite{npjqiKIK}. In the limit $m\to\infty$,
one obtains $K_{\mathrm{mit}}^{(m)}\to U$, indicating that the Taylor
coefficients are bias-free (asymptotically). This result follows directly
from the KIK formula presented in \cite{npjqiKIK} and is also consistent
with the finite-order fidelity analysis developed in the present work.

\subsubsection{$\frac{1}{\sqrt{x}}$ Taylor expansion picture}

It was shown in \cite{npjqiKIK} that even when the noise is not
Hermitian, the following relation holds
\begin{equation}
U=K\frac{1}{\sqrt{K_{I}K}}+O(\Omega_{2}).
\end{equation}
This relation, referred to as the $\emph{KIK formula}$, correctly
captures the small bias arising from the residual second-order Magnus
term $\Omega_{2}$. The formula was later extended to layer-based
noise amplification schemes (``local-folding''). In this framework,
the Taylor coefficients arise from expanding the function $(K_{I}K)^{-1/2}$
around the point $K_{I}K=I$, corresponding to the zero-noise limit.
This provides a rigorous alternative interpretation of the commonly
used zero-noise extrapolation post-processing. From this perspective,
error mitigation reduces to approximating the function $1/\sqrt{x}$
by a power series in $x$. When the noise is Hermitian (see Sec. \ref{subsec: noise Hermitianity}),
the spectrum satisfies $x\in(0,1]$. In practice, however, the noise
spectrum occupies only a subinterval of this range. This observation
allows one to choose a power-series approximation that is optimized
over the effective noise interval, rather than using the Taylor coefficients,
which are optimized at $x=1$. In \cite{npjqiKIK} it was demonstrated
that optimizing over this relevant interval outperforms the Taylor-based
mitigation using the same experimental data. 

In \cite{npjqiKIK}, the effective noise interval was estimated using
an additional echo measurement. In the present work, we introduce
an alternative approach with three key features: (i) it does not require
additional measurements to estimate the noise, (ii) it is automatically
optimized for the specific observable of interest, and (iii) it is
compatible with mid-circuit measurements. The performance and runtime
cost of this approach are analyzed in the remainder of the paper. 

\subsubsection{Drift resilience}

As shown numerically in Supplementary note 3 of \cite{npjqiKIK}
and analytically in the end of Appendix I of \cite{LKIK}, Taylor-based
mitigation is drift resilient provided that agnostic noise amplification
is available and the circuit execution order is chosen correctly.
All noise-amplified circuits must be executed for a small number of
shots during which the noise can be assumed to be stable. This set
of executions is then repeated many times to reach the desired statistical
accuracy. In the extreme case, each set contains a single shot from
each circuit.

\subsubsection{Bias-free}

Since $K_{I}K\cong N^{2}$, the KIK formula yields
\begin{equation}
K_{mit}^{(\infty)}=K\frac{1}{\sqrt{K_{I}K}}\cong UN\frac{1}{\sqrt{N^{2}}}=U.
\end{equation}
$K_{mit}^{(\infty)}$ represents infinite order Taylor mitigation.
As shown in \cite{LKIK}, the small $O(\Omega_{2})$ correction implied
by the $\cong$ sign can be strongly suppressed by layer-wise amplification.
Consequently, the Layered-KIK formula is bias-free for all practical
purposes.

\subsubsection{Hermiticity of the effective noise}\label{subsec: noise Hermitianity}

As explained in \cite{LKIK} and in Appendix \ref{App: hermitianity}
of the present paper, even if the native noise is Hermitian (e.g.
due to decoherence or Pauli errors), the effective noise channel of
the circuit $N$ is generally not Hermitian due to time ordering.
The leading contribution to this non-Hermiticity is given by the second-order
Magnus term $\Omega_{2}$ of $N$. As shown in \cite{LKIK} when
noise amplification is performed in layers, $\Omega_{2}$ for the
full circuit has two contributions: (i) the $\Omega_{2}$ term of
each individual layer, and (ii) cross terms arising from commutators
of the first-order terms $\Omega_{1}$ from different layers.

The first contribution becomes negligible when the number of layers
is sufficiently large (on the order ten). Importantly, this does not
introduce any additional runtime penalty or sampling overhead compared
to single-layer mitigation. The second contribution is eliminated
when Taylor mitigation of order two or higher is applied. Therefore,
when using a sufficiently large number of layers and mitigation order
$M\ge2$, the $\Omega_{2}$ term can be neglected. As a result, the
leading source of non-Hermiticity is canceled when using Layered KIK,
and the effective noise satisfies $N_{\mathrm{eff}}=N_{\mathrm{eff}}^{\dagger}$.
For brevity, we drop the subscript ``eff'' in the remainder of the
paper.

Since the noise is Hermitian, it admits a spectral decomposition in
Liouville space,

\begin{equation}
N=\sum s_{i}\ketbra{s_{i}}{s_{i}},
\end{equation}
where $0<s_{i}\le1$. $s_{i}\to0$ corresponds to infinite noise,
while $s_{i}\to1$ corresponds to zero noise. In this work we assume
the native noise is Hermitian, since it is possible to apply pseudo-twirling
\cite{2024PST} in sufficiently thin layers to render the noise in
each layer Hermitian.

\section{Infidelity and its cost at finite mitigation order }

\begin{figure}
\centering
\includegraphics[width=16cm]{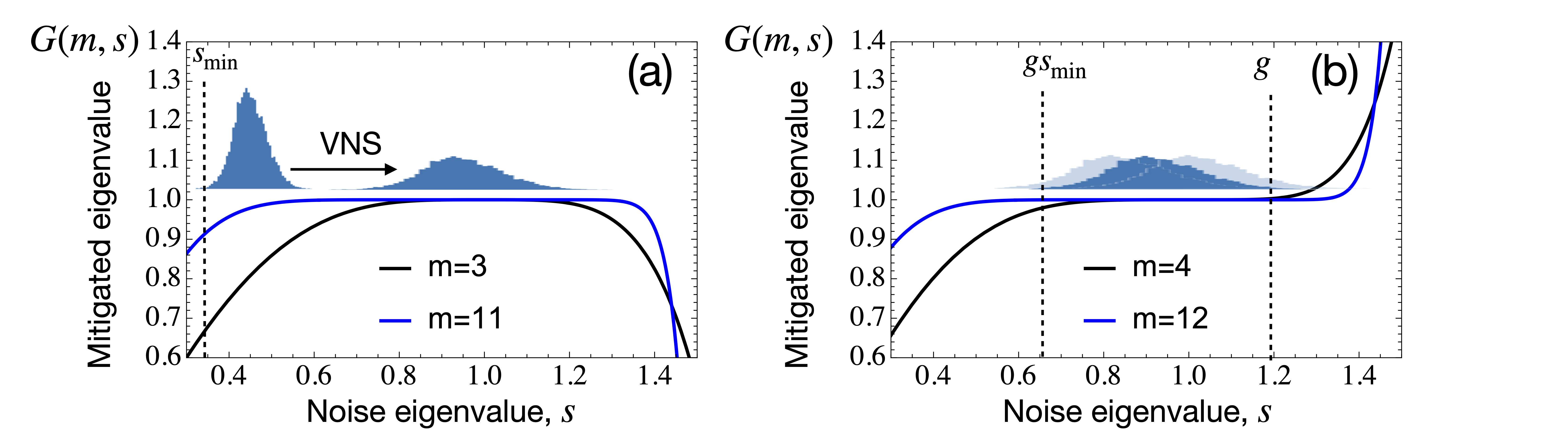}\caption{Plots of the mitigation function $G(m,s)$ for odd (a) and even (b)
orders. As the order increases, the plateaus become wider, and a broader
range of noise eigenvalues is mapped to 1, corresponding to successful
mitigation. The left histogram in (a) represents the noise histogram
of the circuit. By applying our VNS method, the histogram is shifted
to the center of the plateau (right histogram), where the mitigation
performance is significantly improved. The histograms in (b) illustrate
that when the scaled noise interval $[gs_{\min},g]$ is narrower than
the width of the plateau, small changes in $g$ do not affect the
value of the mitigated observable. We exploit this principle to determine
$g$ from the noise-amplified expectation value.}\label{Fig1: G curves}

\end{figure}

Using standard properties of the operator norm $\nrm A_{op}=\sqrt{max[eig(A^{\dagger}A]}$
we get
\begin{equation}
|\left\langle A\right\rangle _{ideal}-\left\langle A\right\rangle _{mit}^{(m)}|\le\nrm{U-K_{mit}^{(m)}}\sqrt{tr\bar{A}^{2}}\sqrt{tr\rho_{0}^{2}},\label{eq: <A> error bound}
\end{equation}
where $\bar{A}$ denotes the observable $A$ with its trace removed.
Using unitary invariance of the operator norm and assuming Hermitian
noise, this bound can be written as

\begin{equation}
\nrm{U-K_{mit}^{(m)}}_{op}=\nrm{I-U^{\dagger}K_{mit}^{(m)}}_{op}=max_{i}|1-\sum_{k=0}^{m}a_{k}^{(m)}s_{i}^{2k+1}|.
\end{equation}
Interestingly, the sum appearing above can be expressed in closed
form as
\begin{equation}
G(m,s_{i})=\sum_{k=0}^{m}a_{k,Tay}^{(m)}s_{i}^{2k+1}=s_{i}\frac{_{2}F_{1}(-m,1/2,3/2,s_{i}^{2})}{_{2}F_{1}(-m,1/2,3/2,1)},
\end{equation}
where $_{2}F_{1}$ is the Gauss hypergeometric function and $_{2}F_{1}(-m,1/2,3/2,1)=\frac{2^{-m}(2m+1)\text{!!}}{m!}=Binomial(m+1/2,m)$.
For convenience, we define $F(m,x)=_{2}F_{1}(-m,1/2,3/2,x)$\@. $G$
is the \textit{mitigation function} and it describes to what extent
the noise eigenvalues are mapped to 1 which corresponds to zero noise.
As shown in Fig. \ref{Fig1: G curves}, larger mitigation orders lead
to broader ``plateaus'' where the input noise eigenvalues are mapped
to one with high accuracy. Previous studies deal with the regime $s\le1$
that contain the physical noise histogram (left histogram in Fig.
\ref{Fig1: G curves} (a)). In this work we exploit the fact that
the plateaus extend into the $s>1$ regime. Our virtual noise scaling
(VNS) technique improves the mitigation performance by shifting the
noise histogram to the center of the plateau (right histogram in Fig.
\ref{Fig1: G curves} (a)).

Since $G(m,s)$ is monotonically increasing in $s$ for any positive
integer $m$ and $0<s<1$, the operator norm is determined by the
smallest singular value $s_{\min}$ and becomes
\begin{equation}
\nrm{U-K_{mit}^{(m)}}_{op}=|1-\sum_{k=0}^{m}a_{k,Tay}^{(m)}s_{\min}^{2k+1}|=1-G(m,s_{min})=1-\frac{s\intop_{0}^{s_{\min}^{2}}(1-t^{2})^{m}dt}{\intop_{0}^{1}(1-t^{2})^{m}dt}\triangleq I_{op}^{(m)}(s_{\min}),\label{eq: Iop}
\end{equation}
where we have used monotonicity and the fact that $\frac{sF(m,s^{2})}{F(m,1)}|_{s=1}=1$
to remove the absolute value. The quantity $I_{op}^{(m)}(s_{\min})$
defines the infidelity function and together with (\ref{eq: <A> error bound})
it quantifies the worst-case error in estimating noiseless expectation
values using $m$-th order Taylor mitigation.

In this work we study the relations between the unmitigated infidelity,
the mitigated fidelity and the runtime overhead (sampling overhead
and depth overhead). The sampling overhead $\gamma_{m}^{2}=(\sum_{k=0}^{m}|a_{k}^{(m)}|)^{2}$
can also be expressed in terms of the Gauss hypergeometric function
\begin{align}
\gamma_{m} & =\frac{F(m,-1)}{F(m,1)}=\frac{\intop_{0}^{1}(1+t^{2})^{m}dt}{\intop_{0}^{1}(1-t^{2})^{m}dt}.
\end{align}
While $\gamma_{m}^{2}$ captures the sampling overhead, the total
runtime overhead $\mathcal{R}$ must also account for the increased
circuit depth associated with noise amplification. An amplification
factor $2j+1$ increases the circuit depth by the same factor, $d=2j+1$.
When this effect is included, the runtime overhead is
\begin{equation}
\mc R=\gamma_{m}^{2}\left\langle d\right\rangle ,
\end{equation}
where $\left\langle d\right\rangle $ is the averaged depth increase.
For $m\ge3$ Taylor mitigation, we find in Appendix \ref{App: depth}
that $\left\langle d\right\rangle \simeq m$. The runtime overhead
grows exponentially with circuit size, since larger mitigation orders
are required to reach a fixed target infidelity. Consequently, QEM
is not a pathway to scalability. Rather, it is a method for obtaining
high-fidelity results from devices whose noise is significant but
not extreme. We refer to this regime of QEM applicability as \textit{benign
noise}, defined by $I_{op}^{(0)}=1-s_{\min}\le1/2$. This is not a
strict definition, and stronger noise can be mitigated at a higher
cost. As hardware improves and QEC becomes practical, the circuit
sizes that fall within the benign-noise regime after QEC will increase.

\begin{figure}
\centering
\includegraphics[width=16cm]{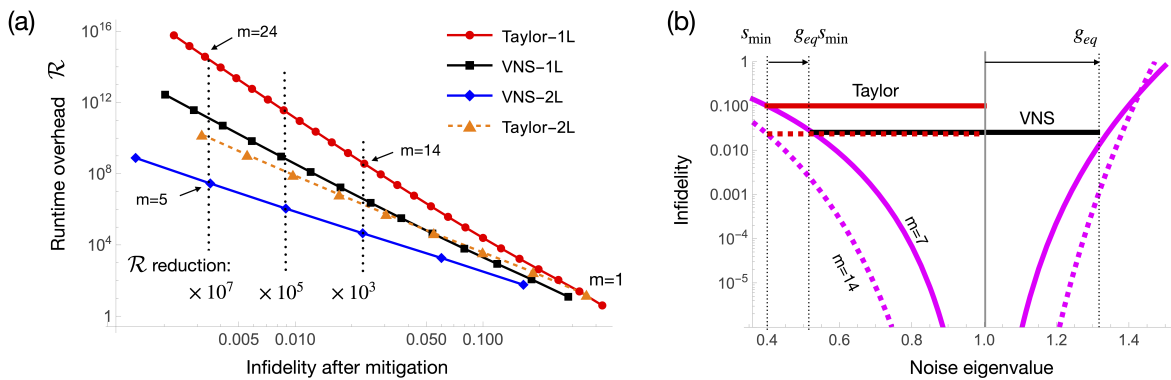}

\caption{(a) The red curve shows the runtime overhead $\protect\mc R$ of
standard Taylor mitigation as a function of the mitigated worst-case
infidelity with the markers indicating the mitigation order $m$.
The black curve, shows that the virtual noise scaling (VNS) introduced
in the present work, leads to a substantial reduction of the runtime
overhead. The initial infidelity is $0.6$ which corresponds to strong
noise. To achieve an even larger reduction in $\protect\mc R$, we
combine VNS with a two-layer mitigation protocol (blue curve). The
corresponding runtime overhead reduction factors for several target
infidelities are indicated below the vertical dotted lines. This orders-of-magnitude
reduction in $\protect\mc R$ can make previously unrealistic mitigation
performance feasible. For reference, the orange curve shows two-layer
mitigation without VNS. (b) The solid purple funnel shows the mitigated
infidelity as a function of the noise eigenvalue for mitigation order
seven ($m=7$). The smallest eigenvalue, $s_{\min}=0.4$, determines
the worst-case infidelity. After applying VNS, the original noise
interval (red) is mapped to the black interval. Although this interval
is wider, the new extreme point $g_{eq}s_{\min}$ corresponds to a
smaller infidelity. For comparison, the dashed purple funnel shows
the infidelity for mitigation order $m=14$, which is the order required
to achieve the same infidelity as the $m=7$ mitigation with VNS.}\label{Fig2: loglog}

\end{figure}

Although $\mathcal{R}$ grows exponentially with $m$, the infidelity
$I_{\mathrm{op}}^{(m)}$ decays approximately exponentially with $m$.
In fact, $\mathcal{R}$ and $I_{\mathrm{op}}$ are related by an approximate
power law. Consequently, $\mathcal{R}(I_{\mathrm{op}})$ appears approximately
linear on a log-log scale. This behavior is illustrated by the red
curve in Fig. \ref{Fig2: loglog}. The initial fidelity in that example
is $I_{\mathrm{op}}^{(0)}=0.6$.

The trade-off between exponential error suppression and exponential
sampling overhead is particularly transparent in the large-$m$ limit,

\begin{align}
I_{op}^{(m)}(s_{\min}) & \underset{m\gg1}{\longrightarrow}\:\frac{(1-s_{\min}^{2})^{m+1}}{\sqrt{\pi m}s_{\min}},\\
\gamma_{m}^{2} & \underset{m\gg1}{\longrightarrow}\:\frac{4^{m+1}}{\pi m}.
\end{align}
For a given initial infidelity $1-s_{\min}$ the linear slope $a=\frac{d\log(\gamma_{m}^{2}m)}{d\log I_{op}^{(m)}(s_{\min})}$
in the log-log implies $\log(\gamma_{m}^{2}m)=a\log I_{op}^{(m)}(s_{\min})+b$
or equivalently $\gamma_{m}^{2}m=e^{b}(I_{op}^{(m)}(s_{\min}))^{a}$,
i.e., a power law. The often-stated exponential overhead with circuit
size is encapsulated in the value of $s_{\min}$. However, for a fixed
circuit, the cost of further reducing the mitigated infidelity follows
a power-law scaling. See \cite{ZNE2025directAnalysis} for an alternative
analysis of ZNE-like mitigation costs. 

\section{VNS - virtual noise scaling}\label{sec: VNS}

The standard Taylor mitigation scheme is based on an expansion around
the point $s=1$. However, since $0<s\leq1$, this expansion point
is generally not centered within the eigenvalue distribution of the
noise, and the resulting approximation is suboptimal. We therefore
propose to rescale all occurrences of the noise operator $N$ by a
constant factor $g$, that is, $K\rightarrow K_{g}=gK$ and $K_{I}\rightarrow K_{I,g}=gK_{I}$.
As a result, the mitigated evolution operator becomes
\begin{align}
K_{mit}^{(m)}(g) & =\sum_{k=0}^{m}a_{k}^{(m)}(g)K(K_{I}K)^{k}\cong U\sum_{k=0}^{m}a_{k}^{(m)}(g)N^{2k+1},
\end{align}
with rescaled coefficients
\begin{equation}
a_{k}^{(m)}(g)=a_{k,Tay}^{(m)}g^{2k+1}.
\end{equation}
Since the eigenvalues of $gN$ are simply $gs_{i}$, one may choose
$g>1$ so that the eigenvalue distribution is shifted closer to the
expansion point $s=1$. Several choices for the scaling factor $g$
are possible, for example,

$g=1/\sqrt{s_{\min}}$, $g=2/(1+s_{\min})$, $g_{\mathrm{eq}}=\sqrt{\frac{2}{s_{\min}^{2}+1}}$
and $g_{det}=1/det(N)^{1/n^{2}}$ where $n$ is the Hilbert space
dimension. The choice $g=g_{\mathrm{det}}$ is equivalent to removing
the trace component of the first Magnus term $\Omega_{1}$, since
$N\cong e^{\Omega_{1}}$. Other choices do not remove the trace but
approximately center $\Omega_{1}$ around zero. While we choose a
specific function $g(s_{\min})$ in order to study the potential of
this approach, we later present a simple procedure for determining
$g$ directly from the measured, noise-amplified expectation values. 

After rescaling by $g$, the spectrum of $gN$ lies in the interval
$[gs_{\min},\,g]$. The corresponding infidelity is therefore
\begin{equation}
\nrm{I-U^{\dagger}K_{mit}^{(m)}(g)}_{op}=\max[I_{op}^{(m)}(gs_{\min}),|I_{op}^{(m)}(g)|]\triangleq I_{op}^{(m)}(s_{\min},g).
\end{equation}
Virtual noise scaling also affects $\gamma_{m}$, which becomes 
\begin{equation}
\gamma_{m}(g)=\sum_{k=0}^{m}\left|a_{k}^{(m)}(g)\right|=\frac{gF(m,-g^{2})}{F(m,1)}
\end{equation}
In the weak-noise limit $s_{\min}\rightarrow1$, one finds that for
$m\geq1$ the infidelity reduction is exponential in $m$, while at
the same time there is no additional sampling overhead,
\begin{align}
\lim_{s_{\min}\to1}\frac{I_{op}^{(m)}(s_{\min},g_{eq})}{I_{op}^{(m)}(s_{\min})} & =2^{-(m+1)},\\
\lim_{s_{\min}\to1}\frac{\gamma_{m}(g_{eq})}{\gamma_{m}(1)} & =1.
\end{align}
However, for $m\gtrsim5$ the infidelity is already below typical
experimental accuracy, and it is therefore not clear that the additional
mitigation power is required. While the advantage of virtual noise
scaling is clear in the strong-noise regime, the associated increase
in runtime overhead cannot be ignored. This motivates a direct comparison
between virtual noise scaling and standard Taylor mitigation at different
orders. In particular, one should compare mitigation orders $m$ and
$m_{\mathrm{VNS}}$ that yield the same infidelity. Although virtual
noise scaling at a given order $m$ is always associated with an increased
$\mc R$, it is possible that a lower-order VNS protocol outperforms
an $m$th-order Taylor mitigation scheme. This behavior is illustrated
in Fig. \ref{Fig2: loglog} (a). The red curve shows the runtime overhead
$\mc R$ of Taylor mitigation as a function of the infidelity for
$s_{\min}=0.4$. The vertical dotted line indicates that, for the
same infidelity, a lower-order VNS protocol achieves a smaller runtime
overhead. In this simulation, we used $g=g_{\mathrm{eq}}$ .

In the large-$m$ limit, the VNS infidelity and $\mc R$ scale as
\begin{align}
I_{op,VNS-1L}^{(m)} & \underset{m\gg1}{\longrightarrow}\:\frac{(1-g_{\mathrm{eq}}^{2}s_{\min}^{2})^{m+1}}{\sqrt{\pi m}g_{\mathrm{eq}}s_{\min}},\\
\gamma_{m}^{2}m & \underset{m\gg1}{\longrightarrow}\:\frac{(1+g_{\mathrm{eq}}^{2})^{2m+2}}{\pi g_{\mathrm{eq}}^{2}}.
\end{align}
Here we used the fact that, for $g=g_{\mathrm{eq}}$, $\max(I_{op}^{(m)}(g_{eq}s_{\min}),|I_{op}^{(m)}(g_{\mathrm{eq}})|)=I_{op}^{(m)}(g_{\mathrm{eq}}s_{\min})$.
For a fixed mitigation order, the infidelity is therefore suppressed
by a factor $\left(1/(1+s_{\min}^{2})\right)^{m+1}$. The VNS advantage
can be visualized more clearly using the infidelity funnels depicted
in \ref{Fig2: loglog}(b). While the VNS improvement is substantial,
achieving orders-of-magnitude reductions in runtime overhead when
the noise is strong requires combining virtual noise scaling with
mitigation in layers, as discussed in Sec. \ref{sec: Mitigation-in-Layers}.
Before doing so, we present a methodology for determining the scaling
factor $g$ without prior knowledge of the noise and without additional
experimental overhead.

\subsection{Determining the virtual noise scaling factor from the measured expectation
values}\label{subsec: g from meas}

While noise-scaling formulas such as $g_{\mathrm{eq}}$ or $\bar{g}$
(see Appendix \ref{App: gbar}) can be useful, they require prior
knowledge of $s_{\min}$. As shown in Appendix \ref{App: sminL2circ},
$s_{\min}$ can be lower bounded by the infidelity contributions of
the individual gates in the circuit. Assuming local noise, and that
each ``gate'' $i$ is associated with a parameter $s_{\min,i}$, we
find $s_{\min}^{\text{circuit}}\ge\prod_{i}s_{\min,i}$. 

Importantly, the term ``gate'' refers to any time interval during
which noise acts. This definition therefore also includes identity
operations on temporarily idle qubits that are subject to noise. If
error correction or error mitigation has been applied prior to our
protocol, $s_{\min,i}$ refers to the smallest eigenvalue after that
operation. In such cases, the product $\prod_{i}s_{\min,i}$ can remain
in the benign regime even for large circuits.

Beyond the possibility of overestimating the noise (that is, underestimating
$s_{\min}$), such bounds do not account for the dependence on the
initial state and the observable of interest. Some initial states
or observables may be insensitive to certain noise mechanisms that
are present in the circuit. As a result, the effective smallest eigenvalue
of the noise, denoted $s_{\min}^{\mathrm{eff}}$, can be larger than
the smallest eigenvalue of the full noise channel. Restricting attention
to noise sources that lie within the effective light cone of the observable,
as done for example in light-cone estimation methods \cite{kim2023evidence,QedmaQesem},
can significantly reduce the sampling overhead.

To address this issue, we propose the following simple procedure.
First, one plots the mitigated expectation value of an observable
$A$ as a function of $g$ 
\begin{equation}
\left\langle A\right\rangle _{mit}^{(m)}(g)=\sum_{k=0}^{m}a_{k}^{(m)}g^{2k+1}\left\langle A\right\rangle _{2k+1},
\end{equation}
where $\bra AK(K_{I}K)^{k}\ket{\rho_{0}}$ is the noise amplified
expectation value of interest. Due to the structure of the infidelity
function $G(m,s)$, choosing $g$ that is too large or too small (below
one) leads to substantial changes in $\langle A\rangle_{\mathrm{mit}}^{(m)}(g)$.
However, for sufficiently large mitigation order $m$, there exists
a ``plateau regime'' in which the infidelity lies below the target
accuracy, as illustrated in Fig. \ref{Fig1: G curves}. If this plateau
is wider than the effective noise width $1-s_{\min}^{\mathrm{eff}}$,
then the scaled noise interval $\{gs_{\min}^{\mathrm{eff}},g\}$ remains
inside the plateau (see the histogram illustrations in \ref{Fig1: G curves}),
and $\langle A\rangle_{\mathrm{mit}}^{(m)}(g)$ becomes largely insensitive
to small variations in $g$ (see piles illustration in \ref{Fig1: G curves}
(b)). This is also illustrated in the infidelity plots in Fig. \ref{Fig2: loglog}(b).

The value of $g$ can therefore be estimated directly from the curve
$\langle A\rangle_{\mathrm{mit}}^{(m)}(g)$ by identifying the range
over which the plateau persists. When the mitigation order is large
and the plateau is wide compared to the effective noise, one may choose
the smallest value $g>1$ for which the deviation from the plateau
center remains within the target accuracy. Since the sampling overhead
increases monotonically with $g$, this choice minimizes the runtime
overhead.

When the mitigation order is too low to exhibit a clear plateau, we
instead search for the closest substitute: either an extremum or,
if no extremum exists in the interval $[1,g_{\max}]$, an inflection
point. For large $m$, it is sufficient to take $g_{\max}=\sqrt{2}$,
since $G(m,g)$ increases rapidly beyond this value (see Fig. \ref{Fig1: G curves}).
For low mitigation orders ($m\sim4$ and below), extrema or inflection
points may appear at values of $g$ larger than $\sqrt{2}$. Since
$G(m,s)$ has an extremum for odd orders and an inflection point for
even orders (see Fig. \ref{Fig1: G curves}), $\langle A\rangle_{\mathrm{mit}}^{(m)}(g)$
often follows the same form. However, since expectation values involve
linear combinations of noise modes $\{s_{i}\}$, an extremum may appear
before the inflection point.

Although more sophisticated algorithms could yield improved accuracy,
we adopt the following simple rule: we first search for an extremum
in the interval $[1,g_{\max}]$, and if none is found, we search for
an inflection point. If neither exists, then either the mitigation
order is too low, in which case such features appear only at higher
order, or the mitigation order is sufficiently high that standard
Taylor mitigation already suppresses the noise below the experimental
accuracy. These two cases are easily distinguished. In the former,
$\langle A\rangle_{\mathrm{mit}}^{(m)}(g)$ varies significantly with
$g$, whereas in the latter a plateau is already present starting
at $g=1$, so we simply choose $g=1$ to minimize the sampling overhead.

\begin{figure}
\centering
\includegraphics[width=16cm]{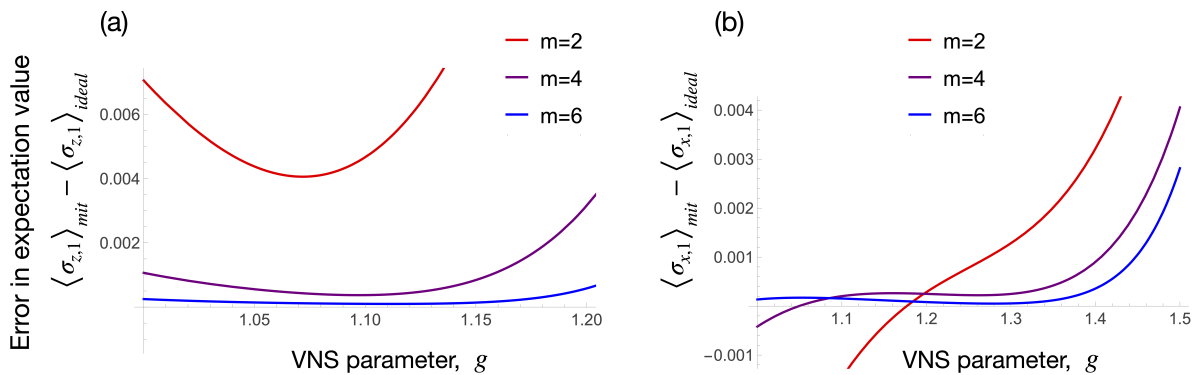}

\caption{(a) The value of g can be extracted from the mitigated expectation
value as a function of the VNS parameter g. As the mitigation order
m increases, a broader plateau develops. The value of $g$ is then
chosen as the extremum or the inflection point within this plateau.
The numerical simulation is based on a four-qubit Ising Trotterized
evolution; see main text for simulation details. In (a), the observable
is $\sigma_{z}$ of the left qubit, whereas in (b) it is $\sigma_{x}$
of the same qubit. While in (a) the optimal value is $g\simeq1.1$,
in (b) it is $g\simeq1.28$. This difference reflects the fact that
the $\sigma_{z}$ observable is less affected by noise (and therefore
requires a smaller value of $g$) and can consequently be mitigated
with a lower runtime overhead.}\label{Fig3: g curves}
\end{figure}

Importantly, the function $\langle A\rangle_{\mathrm{mit}}^{(m)}(g)$
encodes information about both the observable and the initial state.
To illustrate this point, we consider an Ising-like Trotter evolution
with local decoherence noise. Figure \ref{Fig3: g curves}(a) shows
the mitigated expectation value of $\sigma_{z}$ on the left-end qubit
for mitigation orders $m=2,4,$ and $6$. The vertical axis shows
the deviation from the ideal value, and the horizontal axis shows
the VNS parameter $g$. As expected from the discussion above, the
plateau becomes wider as the mitigation order increases, while the
error at the base of the plateau decreases.

Figure \ref{Fig3: g curves}(b) shows the mitigated expectation value
of $\sigma_{x}$ on the same qubit. The initial state and the evolution
are identical to those in Fig. \ref{Fig3: g curves}(a), yet the optimal
scaling parameter is $g\simeq1.27$, compared with $g\simeq1.1$ in
Fig. \ref{Fig3: g curves}(a). This highlights a key advantage of
the present approach: the value of $g$ is not determined by a worst-case
noise estimate, but by the effective manifestation of the noise on
the observable of interest. In this case, $\sigma_{z}$ is less affected
by the noise and therefore exhibits a smaller optimal $g$, which
in turn implies a lower runtime overhead for its mitigation.

This four-qubit simulation contains twenty Trotter evolution steps.
Each step consists of three layers. In the first layer, $R_{zz}(1/30)$
rotations are applied to qubit pairs 1--2 and 3--4. In the second
layer, all qubits undergo an $R_{x}(1/15)$ rotation. Finally, in
the third layer, qubits 2 and 3 undergo an $R_{zz}(1/30)$ rotation.
Local decoherence is applied independently to each qubit. In the first
and third layers, the coefficient of the $\sigma_{z}$ Lindbladians
is $1/200$. During the second layer, the corresponding decay coefficient
is $1/2000$.

\subsection{Explicit form for the first two orders}

For first-order mitigation, no inflection point can occur for $g\neq0$.
Instead, an extremum appears at $g_{extr}^{(1)}=\sqrt{\frac{\left\langle A\right\rangle _{1}}{\left\langle A\right\rangle _{3}}}$
and the corresponding mitigated value is
\begin{equation}
\left\langle A\right\rangle _{mit}^{(1)}=\sqrt{\frac{\left\langle A\right\rangle _{1}^{3}}{\left\langle A\right\rangle _{3}}.}\label{eq: 1st vns explicit}
\end{equation}
This expression is identical to exponential extrapolation with noise
scale factors one and three. This formula should not be used when
$\langle A\rangle_{3}>\langle A\rangle_{1}$ or when $\langle A\rangle_{1}\langle A\rangle_{3}<0$,
as it then yields $g<1$ or an imaginary value of $g$. Such situations
may occur when the ideal expectation value is close to zero and circuit
noise changes the sign of the measured expectation values. We propose
a simple strategy to address this issue. First, choose another observable
$B$ whose expectation value is not close to zero within the experimental
accuracy, that is, an observable that differs from zero by several
standard deviations. Next, compute the mitigated value of $\langle A\rangle$
using
\begin{equation}
\left\langle A\right\rangle _{mit}=\left\langle A+B\right\rangle _{mit}-\left\langle B\right\rangle _{mit},
\end{equation}
where Eq. (\ref{eq: 1st vns explicit}) is separately applied to $A+B$
and $B$. At first order, the effective noise channel may not be Hermitian,
in contrast to the assumption used in higher-order analysis.

For second-order mitigation with virtual noise scaling, an inflection
point occurs at $g_{infl}^{(2)}=\sqrt{\frac{\left\langle A\right\rangle _{3}}{\left\langle A\right\rangle _{5}}}$
and the resulting mitigated value is
\begin{equation}
\left\langle A\right\rangle _{mit}^{(2,infl)}=\frac{15}{8}\sqrt{\frac{\left\langle A\right\rangle _{3}}{\left\langle A\right\rangle _{5}}}\left\langle A\right\rangle _{1}-\frac{7}{8}\sqrt{\frac{\left\langle A\right\rangle _{3}^{5}}{\left\langle A\right\rangle _{5}^{3}}.}\label{eq: 2nd vns explicit}
\end{equation}
As in the first-order case, if $g_{\mathrm{infl}}^{(2)}$ is imaginary,
this indicates a noise-induced sign change of the observable, and
the same strategy described above should be used to address this issue.

\subsection{Virtual noise scaling in dynamic circuits and measurement error mitigation}

In \cite{LKIK} it was shown that amplifying the noise in layers
using the pulse-inverse KIK method leads to two main advantages: (i)
suppression of higher-order Magnus terms to the point where their
residual contribution is negligible for all practical purposes, and
(ii) applicability to dynamic circuits that include mid-circuit measurements
and feedforward. However, the coefficients used in \cite{LKIK} are
still the standard Taylor coefficients. The efficient adaptive coefficients
introduced in \cite{npjqiKIK} rely on knowledge of the effective
noise interval, which is obtained from an echo experiment. However,
this simple echo scheme is incompatible with dynamic circuits, since
the echo is affected not only by gate noise but also by mid-circuit
measurements.

Virtual noise scaling does not suffer from this limitation. VNS can
be interpreted as a layer-dependent shifts of the first Magnus term
$\Omega_{1}$ in each layer, which together represent the noise of
the entire circuit, i.e.,
\begin{align}
gUN & =g_{1}U_{1}e^{\Omega_{1}^{(1)}}\ldots g_{l}U_{l}e^{\Omega_{1}^{(l)}}\ldots g_{L}U_{L}e^{\Omega_{1}^{(L)}}\nonumber \\
 & =U_{1}e^{(\Omega_{1}^{(1)}+\ln g_{1}I)}\ldots U_{l}e^{(\Omega_{1}^{(l)}+\ln g_{l}I)}\ldots U_{L}e^{(\Omega_{1}^{(L)}+\ln g_{L}I)},
\end{align}
and in the same way
\begin{equation}
g^{2j+1}UN^{2j+1}=U_{1}e^{(2j+1)(\Omega_{1}^{(1)}+\ln g_{1}I)}\ldots g_{l}U_{l}e^{(2j+1)(\Omega_{1}^{(l)}+\ln g_{l}I)}\ldots g_{L}U_{L}e^{(2j+1)(\Omega_{1}^{(L)}+\ln g_{L}I)}.
\end{equation}

As a result, the order-by-order cancellation of powers of $\Omega_{1}$
arising from different layers, as established in \cite{LKIK}, remains
valid after the replacement

$\Omega_{1}^{(l)}\rightarrow\Omega_{1}^{(l)}+\ln g_{l}I$. Due to
the virtual noise scaling, the operator norm of $\Omega_{1}^{(l)}+\ln g_{l}I$
is smaller than that of $\Omega_{1}^{(l)}$, which further enhances
the effectiveness of the mitigation.

\begin{figure}
\centering
\includegraphics[width=10cm]{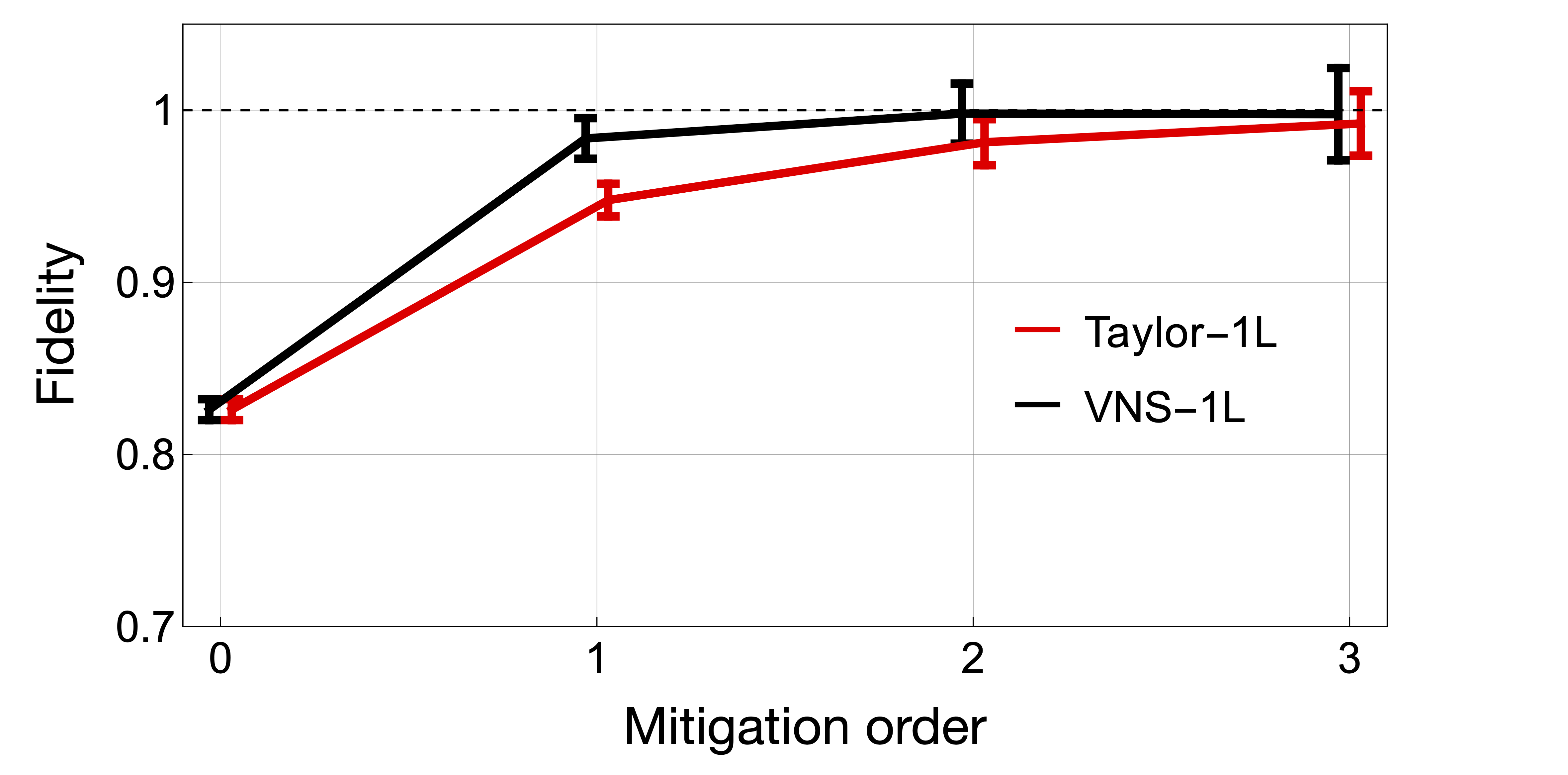}

\caption{The red curve shows the fidelity of a GHZ state creation experiment
reported in \cite{Parity2025REM}. The GHZ state is created using
a dynamic circuit. The mid-circuit measurements, mid-circuit resets,
and final measurements are simultaneously mitigated using the parity
mitigation method, which is resilient to noise drifts. The black curve
shows an improvement over the Taylor post-processing when applying
VNS to the same experimental data. }\label{Fig4: GHZ-fidelity}

\end{figure}

To experimentally demonstrate the applicability of VNS to dynamic
circuits we revisit the four-qubit GHZ dynamic-circuit experiment
carried out in \cite{Parity2025REM}. Figure \ref{Fig4: GHZ-fidelity}
shows how the fidelity curve improves when using VNS on the same experimental
data. The method is also applicable to parity-based or reset-based
SPAM mitigation introduced in \cite{Parity2025REM}. In fact, this
experiment already includes parity-based SPAM mitigation and parity
based mid-circuit reset mitigation. VNS automatically address the
SPAM noise as well. The measurement twirling applied in this experiment
guarantees that the measurement noise is Hermitian. 

\section{Layer-Based Mitigation}\label{sec: Mitigation-in-Layers}

In general, it is possible to divide the circuit into multiple layers,
e.g., A and B, and mitigate each one separately such that
\begin{equation}
K_{mit,\bar{A}\bar{B}}^{(m)}=K_{mit,B}^{(m)}K_{mit,A}^{(m)}.\label{eq: KmitB KmitA}
\end{equation}
In the Appendix \ref{App: depth}, we show that the runtime overhead
in Taylor mitigation becomes 
\begin{equation}
\mc R=\gamma_{tot}^{2}\left\langle d\right\rangle =\gamma_{m}^{2}\gamma_{m}^{2}m.\label{eq: gamma two layers}
\end{equation}
At the same time, the infidelity becomes substantially smaller, since
each layer experiences lower noise levels. For simplicity, we consider
the case where the two layers have the same $s_{\min}$, so that $s_{\min}^{\mathrm{tot}}=s_{\min}^{2}$.
The analysis can be extended to the case $s_{\min}^{B}\neq s_{\min}^{A}$
by using a different mitigation order for each layer. Furthermore,
we note that the layers can be chosen such that $s_{\min}^{A}\simeq s_{\min}^{B}$,
for example by including the same number of two-qubit gates in each
layer. In Appendix \ref{App: sminL2circ}, we show that the unmitigated
two-layer infidelity can be upper bounded by the individual layer
infidelity
\begin{equation}
I_{op,tot}^{(0)}\le I_{op,A}^{(0)}+I_{op,B}^{(0)}-I_{op,A}^{(0)}I_{op,B}^{(0)},
\end{equation}
and for the mitigated infidelity we get
\begin{equation}
I_{op,tot}^{(m)}\le I_{op,A}^{(m)}+I_{op,B}^{(m)}.
\end{equation}
The fact that this bound is less tight, since it does not include
cross terms, is not critical because the mitigated infidelities are
expected to be small, and the product correction is therefore negligible.
Using $s_{\min}^{\mathrm{tot}}=s_{\min}^{2}$ we get
\begin{equation}
I_{op,\bar{A}\bar{B}}^{(m)}\le2I_{op}^{(m)}(\sqrt{s_{\min}^{\text{tot}}}),\label{eq: Tylor 2L inf}
\end{equation}
which is smaller than $I_{op}^{(m)}(s_{\min}^{\text{tot}})$ for any
$m$ and $s_{\min}$. First we compare two-layer mitigation without
VNS, as discussed above, with the standard single-layer Taylor mitigation
(also without VNS). In the $s_{\min}=0.4$ case studied in Fig. \ref{Fig2: loglog},
the dashed orange curve shows that, despite the increased sampling
overhead, two-layer mitigation is beneficial because the improvement
in infidelity is larger. However, this is not true for all values
of $s_{\min}$.

\begin{figure}
\centering
\includegraphics[width=17cm]{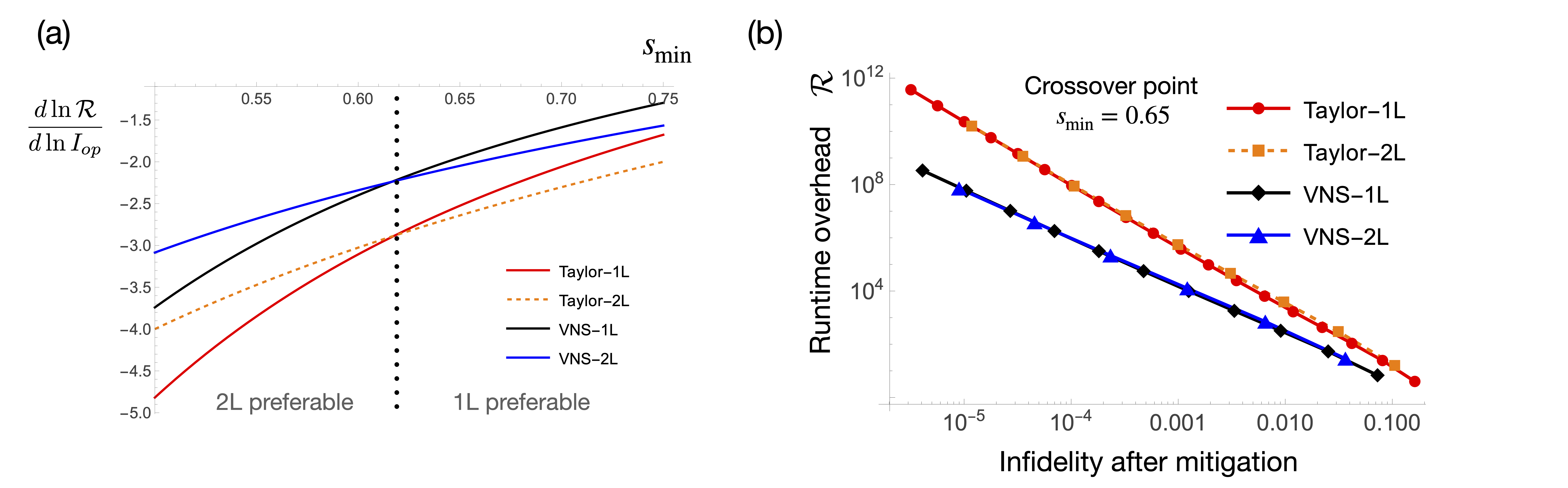}\caption{(a) Slopes of the $\mathcal{R}$-infidelity curves for various mitigation
schemes as a function of the smallest eigenvalue of the noise operator,
$s_{\min}$. Lower absolute values correspond to slower growth in
mitigation overheads as infidelity decreases. Without VNS, two-layer
Taylor mitigation outperforms single-layer mitigation when the noise
is above a critical value ($s_{\min}$ is below a critical value).
The VNS slopes are always less steep (than the Taylor slopes, yet
VNS also shows a crossover: beyond a critical noise value, two-layer
VNS becomes more efficient than single-layer VNS in terms of runtime
overhead. For both Taylor and VNS, the crossover value is $\sim0.62$
for $m\to\infty$, while for finite $m$ we find the crossover to
be $\sim0.65$. (b) Same plot as Fig. \ref{Fig2: loglog}(a) but at
the crossover point $s_{min,2L}^{\text{tot}}=0.65$. While VNS still
outperforms Taylor mitigation at this point, two-layer mitigation
should be used only when the noise is stronger i.e., $s_{min}^{\text{tot}}<s_{min,2L}^{\text{tot}}$. }\label{Fig5: Slopes}
\end{figure}

Using the large-m expansion, we find that the asymptotic (large-m)
slopes of the $\mathcal{R}$-infidelity curves are
\begin{align}
\frac{d\ln\mc R}{d\ln I_{op}}|_{Taylor-1L} & =\frac{2\ln2}{\ln[1-(s_{min}^{\text{tot}})^{2}]},\\
\frac{d\ln\mc R}{d\ln I_{op}}|_{Taylor-2L} & =\frac{4\ln2}{\ln(1-s_{\min}^{\text{tot}})}.
\end{align}
As shown in Fig. \ref{Fig5: Slopes}, a crossover occurs when these
two slopes become equal at $s_{\min}^{\text{tot}}\simeq0.62$. Using
the exact expressions for the infidelity and runtime overhead, we
find that for finite order the crossover instead occurs at $s_{\min,2L}^{\text{tot}}\simeq0.65$.
For $s_{\min}^{\text{tot}}\ge s_{\min,2L}^{\text{tot}}$, corresponding
to weaker noise, single-layer mitigation is preferable. Below this
value, two-layer mitigation is preferable. Note that for $s_{\min}^{\text{tot}}<s_{\min,2L}^{\text{tot}}$,
part of the advantage comes from the fact that, for a given target
infidelity, two-layer mitigation requires a lower mitigation order
and therefore shorter circuits. However, the main contribution comes
from the stronger reduction in infidelity.

For the two-layer case we use an upper bound on the infidelity rather
than the exact infidelity. The exact infidelity depends on the specific
circuit details, while the upper bound does not. As a result, in some
cases the actual infidelity can be substantially smaller than the
upper bound used here.

When further increasing the number of layers, we find an additional
slope crossover between two-layer VNS and three-layer VNS at $s_{\min}\sim0.5$.
However, when the advantage of three-layer VNS becomes significant,
the required runtime overhead is substantial. For example, for I$_{op}^{(0)}=0.75$,
the sampling overhead is $4.7\times10^{8}$ to achieve an infidelity
of $I_{op}^{(m=4)}=0.014.$ Yet, this sampling overhead is $35$ times
smaller than that required for two-layer VNS.

Alternative multi-layer mitigation approaches have been studied in
\cite{Mari2024LayerBasedRich,chen2025LayeredPEC}. However, these
studies rely on different noise inversion schemes and are therefore
outside the scope of the present analysis.

\section{Using both virtual noise scaling and layer-based mitigation}

To further exploit the potential of the VNS approach, we next study
the combination of two-layer mitigation and VNS. With VNS applied
to each layer, and assuming as before that $s_{\min}^{A}\simeq s_{\min}^{B}\simeq\sqrt{s_{\min}^{\text{tot}}}$,
the resulting infidelity is bounded by
\begin{equation}
I_{VNS-2L}^{(m)}\le2I_{op}^{(m)}[g_{eq}(\sqrt{s_{\min}^{\text{tot}}})\sqrt{s_{\min}^{\text{tot}}}],\label{eq: Ivns2L}
\end{equation}
and the two-layer runtime overhead $\mc R$ is
\begin{equation}
\mc R_{VNS-2L}=\gamma^{4}(\sqrt{s_{\min}^{\text{tot}}})m.\label{eq: ROvns2L}
\end{equation}
The advantage of the two-layer VNS (VNS-2L) scheme is illustrated
in Fig. \ref{Fig2: loglog} for $s_{\min}^{\text{tot}}=0.4$. The
blue curve shows that VNS-2L outperforms Taylor-1L, Taylor-2L, and
VNS-1L. As in the no-VNS case discussed in the previous section, we
observe a crossover between VNS-2L and VNS-1L. VNS-2L has lower runtime
overhead compared to VNS-1L when $s_{\min}^{\text{tot}}\le s_{\min,2L}\simeq0.65$.
Figure \ref{Fig5: Slopes} shows the crossover in the large $m$ limit.

As shown in Fig. \ref{Fig2: loglog}, for a mitigated infidelity of
0.024 (right vertical line), we find a reduction of $\mc R$ from
$3.6\times10^{8}$ in the Taylor-1L method to $3.8\times10^{4}$ in
the VNS-2L method. For lower mitigated infidelity, the advantage is
even larger. If the mitigated infidelity is $9\times10^{-3}$, we
obtain a reduction of $\mc R$ from $3.6\times10^{11}$ in the Taylor-1L
method to $9.4\times10^{5}$ in the VNS-2L method. This corresponds
to a reduction of $\mc R$ by a factor of $10^{4}$ in the first case
and $10^{5}$ in the second case. The asymptotic slopes ($m\gg1$)
in the VNS case are
\begin{align}
\frac{d\ln\mc R}{d\ln I_{op}}|_{VNS-1L} & =\frac{\ln\left(\frac{2}{(s_{\min}^{\text{tot}})^{2}+1}+1\right)}{\ln\left(\frac{2}{(s_{\min}^{\text{tot}})^{2}+1}-1\right)},\\
\frac{d\ln\mc R}{d\ln I_{op}}|_{VNS-2L} & =\frac{2\ln\left(\frac{2}{s_{\min}^{\text{tot}}+1}+1\right)}{\ln\left(\frac{2}{s_{\min}^{\text{tot}}+1}-1\right)}.\label{eq: vns2Lslope}
\end{align}
Equations (\ref{eq: Tylor 2L inf}) - (\ref{eq: vns2Lslope}) together
with Figs. \ref{Fig2: loglog} and \ref{Fig5: Slopes} constitute
our main findings: orders of magnitude reduction compared to the baseline
one-layer Taylor mitigation (Richardson extrapolation). 

\section{Extension to General Constant-Step Amplification Factors}

In the preceding sections we have considered amplification factors
of the form $\{1+2j\}_{j=0}^{m}.$ However, other sets of amplification
factors have been investigated in the literature \cite{Tiron,fracZNEvolume,majumdar2023best,kim2023evidence},
motivating an extension of our analysis to general constant-step amplification
sets of the form $\{1+j\alpha\}_{j=0}^{m}$ where $\alpha>0$ is the
amplification step.

\subsection{Analytical Framework for Arbitrary $\alpha>0$}

Presently, agnostic noise amplification (ANA) is restricted to even
values of $\alpha$ due to the discrete nature of unitary folding
schemes. However, characterization-based techniques such as probabilistic
error amplification (PEA) \cite{kim2023evidence} can generate fractional
amplification steps as well as non-uniform steps where $\alpha$ depends
on $j$. In analogy to the results derived in Sec. \ref{sec: VNS},
we find that for any $\alpha>0$, the mitigation function $G(m,s_{i},\alpha)$
can be expressed in terms of the Gauss hypergeometric function:
\begin{equation}
G(m,s_{i},\alpha)=\sum_{j=0}^{m}a_{j}^{(m,\alpha)}s_{i}^{1+j\alpha}=s_{i}\left(\begin{array}{c}
m+1/\alpha\\
m
\end{array}\right)\,_{2}F_{1}(-m,\frac{1}{\alpha},1+\frac{1}{\alpha},s_{i}^{d}),
\end{equation}
where $\left(\begin{array}{c}
m+1/\alpha\\
m
\end{array}\right)$ is the binomial coefficient and
\begin{equation}
a_{j}^{(m,\alpha)}=\frac{(-1)^{j}}{j\alpha+1}\binom{m+1/\alpha}{m}\binom{m}{j}.
\end{equation}
The resulting mitigated expectation value of an observable $A$ using
VNS is:
\begin{equation}
\langle A\rangle_{mit}^{(m,\alpha)}=\sum_{j=0}^{m}a_{j}^{(m,\alpha)}g^{1+j\alpha}\langle A\rangle_{(1+j\alpha)},
\end{equation}
where $\langle A\rangle_{(1+j\alpha)}$ is the expectation value measured
with noise amplified by a factor of $1+j\alpha$. The operator norm
infidelity is
\begin{equation}
I_{op}^{(m)}(gs_{min})=\max\left[1-G(m,gs_{min},\alpha),|1-G(m,g,\alpha)|\right].
\end{equation}
If $s_{min}$ is known or estimated, the optimal VNS parameter $g$
in the large $m$ limit is:
\begin{equation}
g=\left(\frac{2}{1+s_{min}^{\alpha}}\right)^{1/\alpha}.
\end{equation}
The corresponding sampling cost $\gamma^{2}$ is:
\begin{equation}
\gamma^{2}=\left[g\binom{m+1/\alpha}{m}{}_{2}F_{1}\left(-m,\frac{1}{\alpha},1+\frac{1}{\alpha},-g^{\alpha}\right)\right]^{2}.
\end{equation}

\subsection{Performance for $\alpha=2n$}

For even values of $\alpha$, ANA can be employed to ensure resilience
against temporal noise drifts. We compared the performance of $\alpha=2$
case studied in the previous sections to $\alpha=4$ and $\alpha=6$.
In the single-layer case, if circuit depth costs are ignored, $\alpha=4$
and $\alpha=6$ offer a mild advantage over $\alpha=2$ in the weak-noise
regime ($s_{min}\simeq0.95$). However, this advantage vanishes once
the increased length of the $\alpha=4,6$ circuits is accounted for.
Similarly, we studied two-layer mitigation with VNS for $\alpha=4$
and $\alpha=6$. While a marginal advantage exists for $s_{min}\gtrsim0.9$,
we previously established that above the $s_{min}=0.63$ threshold,
a single-layer VNS outperforms a two-layer VNS. Consequently, we conclude
that $\alpha=4$ and $\alpha=6$ based on ANA offer no cost-performance
advantage over the standard $\alpha=2$ case.

\subsection{Fractional $\alpha$ and Instability for small $\alpha$.}

\begin{figure}
\centering
\includegraphics[width=17cm]{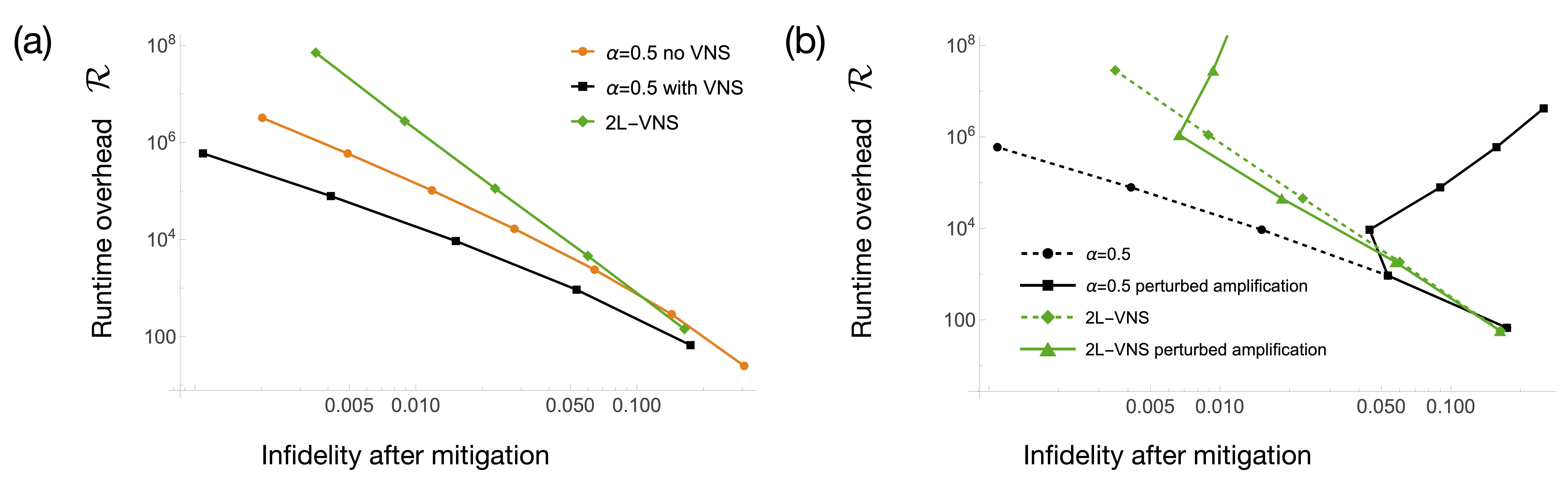}

\caption{(a) Comparison of single-layer mitigation with $\alpha=0.5$ using
VNS (black) and without VNS (orange) to the two-layer $\alpha=2$
VNS protocol (green) studied in Fig. \ref{Fig2: loglog} ($s_{min}=0.4$).
The $\alpha=0.5$ protocol clearly outperforms the 2L-VNS ($\alpha=2$)
case under ideal amplification. (b) The mitigation performance (solid-black)
is severely degraded for $\alpha=0.5$ when the nominal amplification
factor 1.5 is changed to 1.505 without adjusting the other amplification
factors. By contrast, the $\alpha=2$ protocol is much less sensitive
to such deviations. The dashed line show the performance without the
perturbation.}\label{fig: alpha}
\end{figure}

Next, we investigated fractional amplification factors. Unlike ANA,
PEA ensures all circuits have the same physical length, meaning the
runtime overhead is determined solely by $\gamma^{2}$. As shown in
Fig. \ref{fig: alpha}a, single-layer $\alpha=0.5$ mitigation with
VNS (solid black curve) outperforms the two-layer $\alpha=2$ VNS
studied in previous sections. However, this performance comes at the
cost of stability. When $\alpha$ is small, the mitigation becomes
highly sensitive to calibration errors. 

As illustrated in Fig. \ref{fig: alpha}b, modifying the nominal amplification
factor $1.5$ to $1.505$, while keeping all mitigation coefficients
unchanged, completely degrades the $\alpha=1/2$ mitigation performance
with respect to the unmodified amplification shown in the dashed curve.
In contrast, the $\alpha=2$ case (two-layer VNS) is significantly
more robust to such deviations. This heighten sensitivity at $\alpha=0.5$
arises from the larger amplitude of the $a_{j}^{(m,\alpha)}$ coefficients
at small $\alpha$.

We therefore conclude that, in practice, choosing $\alpha=0.5$ (or
smaller values of $\alpha$) is advantageous only if noise amplification
can be tightly controlled. Achieving such control may be challenging
in the presence of non-negligible temporal noise drifts.

\section{Concluding remarks}

The analysis based on the VNS scaling factor $g=\sqrt{2/(1+s_{\min}^{2})}$
is agnostic to the target observable, initial condition, circuit size,
noise type, and circuit topology. As a result, the conclusions are
broadly applicable. Regarding the noise model, we assume that the
native noise is Hermitian and discuss how arbitrary Markovian noise
can be brought into this form. The results presented here correspond
to the worst-case scenario. In non--worst-case situations, VNS may
provide little or no benefit. However, this is not due to a limitation
of VNS, but rather to the exceptionally strong performance of Taylor
mitigation as a starting point. Thus, even in such cases, the observable
error remains bounded by the worst-case VNS error limits derived in
this work.

While $s_{\min}^{2}$ (or more precisely, a lower bound on it) can
be estimated from known gate errors, we also introduce a method that
accounts for the initial condition and the target observable. In this
approach, the VNS parameter $g$ is determined by analyzing the extrema
and inflection points of the mitigated expectation value as a function
of $g$. This can significantly reduce the runtime overhead compared
to the worst-case estimate.

VNS is compatible with dynamic circuit measurements and inherits the
drift-resilience and bias-free properties of the Layered-KIK method
\cite{LKIK}. It can also be used to improve the performance of noise-amplification-based
measurement error mitigation \cite{Parity2025REM}. VNS can also
enhance the performance of probabilistic error amplification schemes,
including cases where the amplification factors are fractional. Our
analysis shows that although small amplification increments can be
efficient, they exhibit an intrinsic instability when the implemented
amplification factors deviate even slightly from their intended values.

The relative reduction in the runtime overhead required to reach a
target infidelity is substantial in the strong-noise regime. However,
this improvement may not be immediately visible on current devices,
since the most pronounced gains occur when the baseline runtime overhead
without VNS is already extremely large. For example, improvements
of $10^{4}$ and 1$0^{5}$ were observed for baseline overheads of
$10^{8}$ and $10^{11}$, respectively. Even after these improvements,
the remaining runtime overhead is still significant. Consequently,
in addition to standard sequential execution, massive parallel execution
of shots is required across different qubit sets, quantum cores, and
even multiple quantum computers. VNS transforms previously inconceivable
overheads into challenging but realistic targets, especially as large-scale
production of NISQ and fault-tolerant devices becomes feasible.
\begin{acknowledgments}
Raam Uzdin is grateful for support from the Israel Science Foundation
(Grants No. 2724/24). The support of the Israel Innovation Authority
is greatly appreciated.
\end{acknowledgments}

\bibliographystyle{apsrev4-1}
\bibliography{Refs_VNS}

\section*{Appendix I - Effective Hermiticity of the noise}\label{App: hermitianity}

In general, the noise in the circuit is not Hermitian, even if the
native Lindbladian noise generators $\mc L(t)$ are Hermitian. Using
the Magnus expansion in the interaction picture in Liouville space,
the noise channel of the full circuit is
\begin{equation}
N=e^{\Omega_{1}+\Omega_{2}+O(\Omega_{3})},
\end{equation}
where
\begin{align}
\Omega_{1} & =\intop^{t}\mc L^{int}(t')dt',\\
\Omega_{2} & =\frac{1}{2}[\intop^{t}\intop^{t'}dtdt'[\mc L^{int}(t'),\mc L^{int}(t'')],
\end{align}
and 
\begin{equation}
\mc L^{int}(t)=\mc U^{\dagger}(t)\mc L(t)\mc U(t).
\end{equation}

Next, we assume the native noise is either inherently Hermitian or
is rendered Hermitian by pseudo-twirling in sufficiently small layers.
If $\mc L^{\dagger}(t)=\mc L(t)$ then $\Omega_{1}=\Omega_{1}^{\dagger}$,
$\Omega_{2}=-\Omega_{2}^{\dagger}$ and $\Omega_{3}=\Omega_{3}^{\dagger}$.
Thus, the leading-order correction to Hermiticity is $\Omega_{2}$.
In \cite{LKIK} , it was shown that amplification by layers (not
to be confused with mitigation by layers in the current paper) leads
to strong suppression of $\Omega_{2}$. When the layers are sufficiently
thin, this suppression scales as $1/N_{\mathrm{layers}}^{2}$, where
$N_{\mathrm{layers}}$ is the chosen number of layers. Note that the
layers are not necessarily related to gates or groups of gates; a
single gate can be sliced into multiple layers. When the layers are
not sufficiently thin, there is a bound showing that the $\Omega_{2}$
contribution decreases as $1/N_{\mathrm{layers}}$.

The $\Omega_{2}$ contribution is typically much smaller than $\Omega_{1}$
when the noise is \textit{benign} (see definition in the main text).
Therefore, dozen layers often make the effective $\Omega_{2}$ contribution
negligible. In this case, the effective noise channel is
\begin{equation}
N_{\mathrm{eff}}=N_{\mathrm{eff}}^{\dagger}+O(\Omega_{2}^{3},\Omega_{4}).
\end{equation}
Note that $\Omega_{2}^{2}$ is Hermitian when $\mathcal{L}^{\dagger}(t)=\mathcal{L}(t),$
so potential corrections arise only from $\Omega_{2}^{3}$. We conclude
that when using the Taylor coefficients (or the scaled Taylor coefficients
studied in this work) together with amplification in layers, the effective
noise circuit is Hermitian for all practical purposes if $\mathcal{L}^{\dagger}(t)=\mathcal{L}(t)$.

\section*{Appendix II - Average circuit depth}\label{App: depth}

\begin{figure}
\centering
\includegraphics[width=10cm]{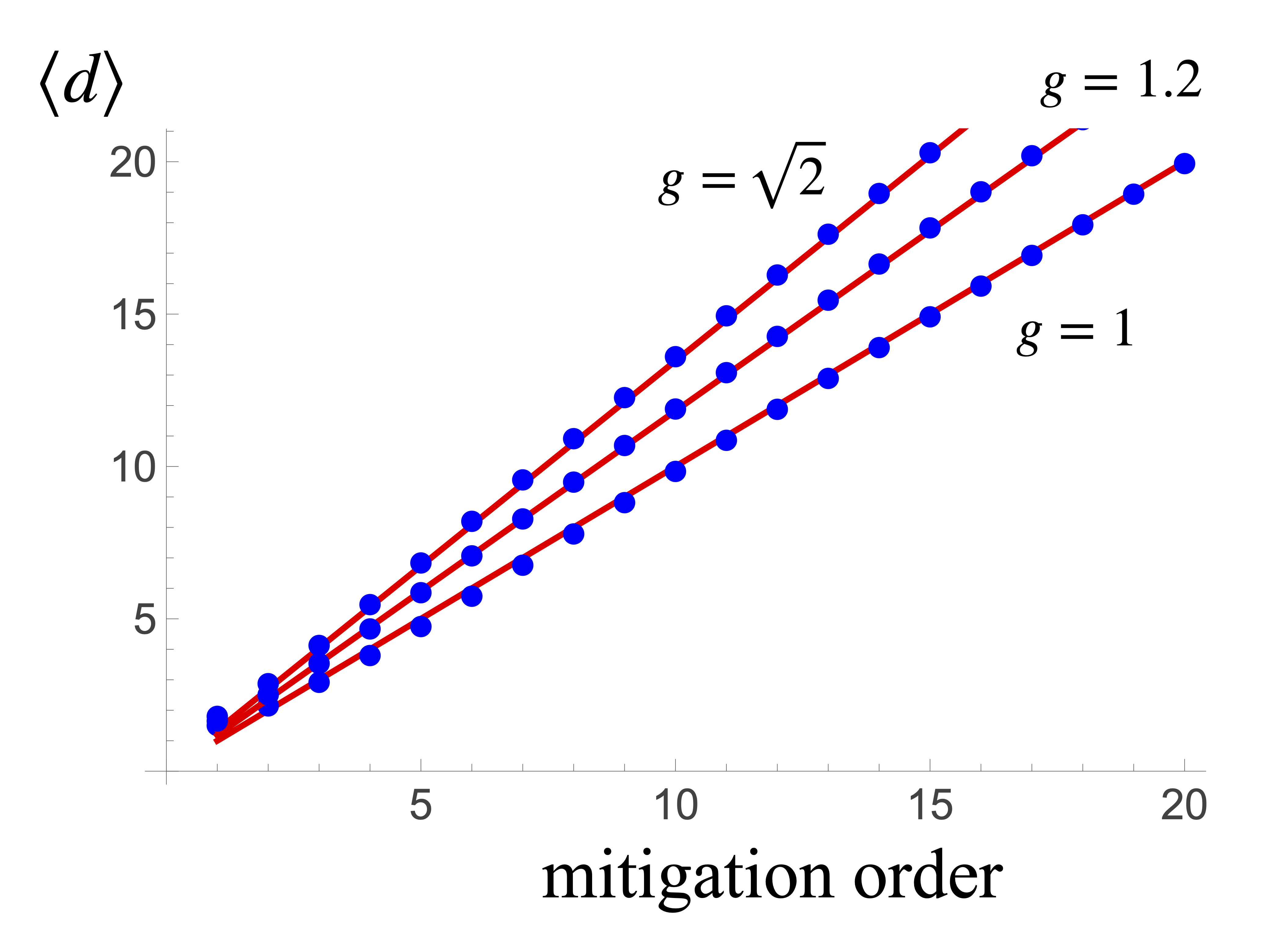}

\caption{Average circuit depth $\left\langle d\right\rangle $ as a function
of the mitigation order $m$ for different values of the VNS parameter
$g$. Markers show the numerical values, while the lines represent
the approximation $\left\langle d\right\rangle \simeq(1+\ln g)m$.}\label{Fig5: depth}
\end{figure}

For a single layer, we optimize the shot allocation such that, for
a fixed total number of shots $N_{tot}=\sum N_{i}$, the variance
$\sum_{i}\frac{a_{i}^{2}}{N_{i}}$ is minimized. This minimization
yields $\sum_{i}\frac{a_{i}^{2}}{N_{i}}=\frac{1}{N_{tot}}(\sum_{i}|a_{i}|)^{2}$.
This result can be obtained using a Lagrange multiplier. Setting the
partial derivatives to zero we get 
\begin{equation}
\frac{d}{dN_{i}}(\sum_{i=0}^{m}\frac{a_{i}^{2}}{N_{i}}+\lambda(\sum_{i=0}^{m}N_{i}-N_{tot}))=-\frac{a_{i}^{2}}{N_{i}^{2}}+\lambda=0,
\end{equation}
we find
\begin{align}
N_{i} & =\left|a_{i}\right|/\lambda,\\
\sum_{i=0}^{m}N_{i} & =\sum_{i=0}^{M}\left|a_{i}\right|/\lambda,\\
N_{tot}= & \gamma/\lambda,\\
a_{i}^{2}/N_{i} & =\left|a_{i}\right|\lambda=\left|a_{i}\right|\gamma/N_{tot},
\end{align}
Where $\gamma\doteq\sum_{i=0}^{M}\left|a_{i}\right|$. As a result,
the variance of the average over $N_{tot}$ shots is
\begin{equation}
\sum_{i=0}^{m}\frac{a_{i}^{2}}{N_{i}}=\gamma^{2}/N_{tot}.
\end{equation}
Next, we assume that $N_{tot}$ already includes the sampling overhead
$\gamma^{2}$, and we calculate the total runtime while accounting
for the fact that different noise-amplified circuits have different
depths. We find
\begin{align}
N_{tot}\sum_{i=0}^{m}\frac{N_{i}}{N_{tot}}(2i+1) & =N_{tot}\sum_{i=0}^{m}\frac{|a_{i}|/\lambda}{N_{tot}}(2i+1)=N_{tot}\sum_{i=0}^{m}\frac{|a_{i}|N_{tot}/\gamma}{N_{tot}}(2i+1)=N_{tot}\sum_{i=0}^{m}\frac{|a_{i}|}{\gamma}(2i+1).
\end{align}
Hence, the average circuit depth is
\begin{equation}
\left\langle d\right\rangle =\sum_{i=0}^{m}\frac{|a_{i}|}{\gamma}(2i+1).
\end{equation}
The runtime overhead is given by the time required to run $N_{tot}$
shots with the averaged increased circuit depth, divided by the time
required to run $N$ shots of the unamplified circuit with the same
accuracy. Using $N_{tot}=\gamma^{2}N$ , we obtain that the runtime
overhead is
\begin{equation}
\mc R=\gamma^{2}\left\langle d\right\rangle .
\end{equation}
As shown in Fig. \ref{Fig5: depth}, $\left\langle d\right\rangle $
is predominantly linear in the mitigation order $m$. In particular
we find that
\begin{equation}
\left\langle d\right\rangle =\sum_{i=0}^{m}\frac{|a_{i}(g)|}{\gamma}(2i+1)\cong(1+\ln g)m.\label{eq: d approx}
\end{equation}

\subsection*{Two-layer case}

Assuming the first layer occupies a fraction $\alpha$ of the total
circuit length, we obtain

\begin{align}
\left\langle d\right\rangle =\sum_{i,j=0}^{m}\frac{|a_{i}|}{\gamma}\frac{|a_{j}|}{\gamma}[\alpha(2i+1)+(1-\alpha)(2j+1)] & =\sum_{i,j=0}^{m}\frac{|a_{i}|}{\gamma}\frac{|a_{j}|}{\gamma}(2i+1)\alpha+\sum_{i,j=0}^{m}\frac{|a_{i}|}{\gamma}\frac{|a_{j}|}{\gamma}(2j+1)(1-\alpha)\nonumber \\
 & =\sum_{i,j=0}^{m}\frac{|a_{i}|}{\gamma}(2i+1)=(1+\ln g)m,
\end{align}
where we have used approximation in Eq. (\ref{eq: d approx}).

\section*{Appendix III - Analytical choice of the VNS parameter}\label{App: gbar}

We start with the large m approximation for the infidelity
\begin{equation}
I_{op}^{(m\gg1)}(s_{\min})\simeq\frac{\left(1-s_{\min}^{2}\right)^{m+1}}{\sqrt{\pi m}s_{\min}}.
\end{equation}
Next, to obtain comparable errors at both ends of the scaled noise
interval $[gs_{\min},g]$, we impose the condition:
\begin{equation}
I_{op}^{(m\gg1)}(gs_{\min})=I_{op}^{(m\gg1)}(g).
\end{equation}
This simplifies to
\begin{equation}
\frac{\left(1-g^{2}s_{\min}^{2}\right)^{m+1}}{s_{\min}}=\left(g^{2}-1\right)^{m+1},\label{eq: derive gbar}
\end{equation}
and finally yields 
\begin{equation}
\bar{g}=\sqrt{\frac{1+s_{\min}^{\frac{1}{m+1}}}{s_{\min}^{2}+s_{\min}^{\frac{1}{m+1}}}}.
\end{equation}
Since $0<s_{\min}\le1$, we get that
\begin{equation}
\bar{g}=\sqrt{\frac{1+s_{\min}^{\frac{1}{m+1}}}{s_{\min}^{2}+s_{\min}^{\frac{1}{m+1}}}}\ge\sqrt{\frac{2}{s_{\min}^{2}+1}}=g_{eq}.
\end{equation}
We have not observed a significant advantage in using $\bar{g}$ instead
of $g_{eq}$. For low orders, $\bar{g}$ slightly overestimates the
$g_{exact}$ value for which $I_{op}(g_{exact}s_{\min})=I_{op}(g_{exact})$.
Hence, the actual infidelity value is $\max(I_{op}(\bar{g}s_{\min},I_{op}(\bar{g}))=I_{op}(\bar{g})$.
In contrast, $g_{eq}$ slightly underestimate $g_{exact}$, and therefore
$\max(I_{op}(g_{eq}s_{\min}),I_{op}(g_{eq}))=I_{op}(g_{eq}s_{\min})$.
Note that $g_{eq}$ can be obtained directly from Eq. (\ref{eq: derive gbar})
by setting $s_{\min}=1$ in the denominator of the expression in the
left-hand side. Alternatively, it holds that $g_{eq}=\lim_{m\to\infty}\bar{g}$. 

\section*{Appendix IV - Relating infidelities of layers to circuit infidelity}\label{App: sminL2circ}

We start by relating the smallest singular value of the whole circuit
to that of the layers in the absence of mitigation. For simplicity
we consider two layers
\begin{equation}
K_{\ensuremath{AB}}=U_{B}N_{B}U_{A}N_{A}\doteq U_{B}U_{A}\tilde{N}_{B}N_{A}.
\end{equation}
Since the infidelity is determined by $I_{op}^{(m)}(s_{\min})$ we
focus first on the smallest singular value $s_{\min}$
\begin{equation}
s_{\min}=\frac{1}{\nrm{N^{-1}}_{(op)}},
\end{equation}
and therefore
\begin{equation}
s_{\min}(K_{\ensuremath{AB}})=\frac{1}{\nrm{(\tilde{N}_{B}N_{A})^{-1}}}=\frac{1}{\nrm{N_{A}^{-1}\tilde{N}_{B}^{-1}}}\ge\frac{1}{\nrm{N_{A}^{-1}}}\frac{1}{\nrm{N_{B}^{-1}}}=s_{\min}(K_{A})s_{\min}(K_{B}),
\end{equation}
where we have used unitary invariance and submultiplicativity of the
operator norm. From this one concludes that for multiple layers the
total $s_{\min}$ can be estimated by knowing the noise of the layers
(or the individual gates). The estimate will overestimate the error
due to the inequalities used in the derivation. More generally we
can write:
\begin{equation}
s_{\min}(K^{tot})\ge\prod_{l}s_{\min,l}.\label{eq: prod est smin}
\end{equation}
First we address the infidelity of the unmitigated circuits $I_{op}^{(0)}=1-s_{\min}$
(for Hermitian noise). Using the definition of $I_{op}^{(0)}$ in
(\ref{eq: prod est smin}) leads to
\begin{equation}
I_{op,tot}^{(0)}=1-s_{\min}^{tot}\le1-\prod_{l}(1-I_{op,l}^{(0)}).\label{eq: inf ord 0 prod}
\end{equation}
This relation sets an upper bound on the infidelity of the whole circuit
in terms of the layers infidelity. This bound is more refined than
simply summing the infidelities. Yet, at the same time, it is still
a simple function of the layer infidelities. For example, in two layers
we have
\begin{equation}
I_{op,tot}^{(0)}\le I_{op,A}^{(0)}+I_{op,B}^{(0)}-I_{op,A}^{(0)}I_{op,B}^{(0)}.\label{eq: two layer ord0 inf}
\end{equation}
Next, we want to treat the total infidelity in mitigation of two layers
where each layer is potentially mitigated to a different order:
\begin{equation}
K_{\ensuremath{mit},tot}^{(m_{A},m_{B})}=K_{\ensuremath{mit},B}^{(m_{B})}K_{\ensuremath{mit},A}^{(m_{A})}.
\end{equation}
The infidelity for Hermitian noise is
\[
\nrm{U_{B}U_{A}-K_{\ensuremath{mit},tot}^{(m_{A},m_{B})}}=
\]
\begin{align}
\nrm{I-U_{A}^{\dagger}[I-(I-\sum_{s_{B}}f_{m_{B}}(s_{B})\ketbra{s_{B}}{s_{B}})]U_{A}[I-(I-\sum_{s_{A}}f_{m_{A}}(s_{A})\ketbra{s_{A}}{s_{A}})]} & =\nonumber \\
\left\Vert U_{A}^{\dagger}(I-\sum_{s_{B}}f_{m_{B}}(s_{B})\ketbra{s_{B}}{s_{B}})U_{A}+(I-\sum_{s_{A}}f_{m_{A}}(s_{A})\ketbra{s_{A}}{s_{A}})\right. & -\\
\left.-U_{A}^{\dagger}(I-\sum_{s_{B}}f_{m_{B}}(s_{B})\ketbra{s_{B}}{s_{B}})U_{A}(I-\sum_{s_{A}}f_{m_{A}}(s_{A})\ketbra{s_{A}}{s_{A}})\right\Vert .
\end{align}
By using the triangle inequality and submultiplicativity for the last
term we get
\begin{equation}
\nrm{U_{B}U_{A}-K_{\ensuremath{mit},tot}^{(m_{A},m_{B})}}\le I_{op}^{(m_{A})}(s_{\min,A})+I_{op}^{(m_{B})}(s_{\min,B})+I_{op}^{(m_{A})}(s_{\min,A})I_{op}^{(m_{B})}(s_{\min,B}).\label{eq: inf prod bound any ord}
\end{equation}
While for $m_{A}=m_{B}=0$ the bound (\ref{eq: two layer ord0 inf})
is tighter than this bound, the current bound is applicable to any
$m_{A},m_{B}$. Moreover, when $m_{A}$ and $m_{B}$ are sufficiently
large $I_{op}^{(m_{A})}(s_{\min,A})+I_{op}^{(m_{B})}(s_{\min,B})\gg I_{op}^{(m_{A})}(s_{\min,A})I_{op}^{(m_{B})}(s_{\min,B})$
and the product term in (\ref{eq: inf prod bound any ord}) can be
neglected.
\end{document}